\documentclass[prl,aps,twocolumn,amsmath,amssymb,floatfix,
superscriptaddress]{revtex4-1}
\usepackage[dvips]{graphics}
\usepackage{color}
\definecolor{dred}{rgb}{0,0,0.6}
\usepackage{graphicx}  
\usepackage{dcolumn}   
\usepackage{bm}        
\usepackage{amssymb}   
\usepackage{amsmath}
\usepackage{braket}
\usepackage[mathscr]{euscript}
\usepackage{color}
\usepackage{hyperref}
\begin{document}

\title{Electrical analogue of one-dimensional and quasi-one-dimensional Aubry-Andr\'{e}-Harper lattices}

\author{Sudin Ganguly}

\email{sudinganguly@gmail.com}

\affiliation{Department of Physics, School of Applied Sciences, University of Science and Technology Meghalaya, Ri-Bhoi-793101, India}

\author{Santanu K. Maiti}

\email{santanu.maiti@isical.ac.in}

\affiliation{Physics and Applied Mathematics Unit, Indian Statistical
  Institute, 203 Barrackpore Trunk Road, Kolkata-700108, India}

\date{\today}

\begin{abstract}
The present work discusses the possibility to realize correlated disorder in electrical circuits and studies the localization phenomena in terms of two-port impedance. The correlated disorder is incorporated using the Aubry-Andr\'{e}-Harper (AAH) model. One-dimensional and 
quasi-one-dimensional AAH structures are explored and directly mapped with their tight-binding analogues. Transitions from the high-conducting phase to the low-conducting one are observed for the circuits.
\end{abstract}


\maketitle

The Aubry-Andr\'{e}-Harper (AAH) model~\cite{aah1,aah2} represents a classic example of a one-dimensional (1D) quasi-crystal, possesses several intriguing features. In the nearest-neighbor tight-binding (TB) framework, the 1D AAH model with an incommensurate potential exhibits a sharp localization-delocalization transition, where all the eigenstates are delocalized below a critical point, while all of them are completely localized beyond that critical point. Due to the incommensurate potential, this 1D model shows a gapped and fractal-like energy spectrum. Beyond the minimal nearest-neighbor TB model, energy-dependent mobility edges have also been predicted analytically~\cite{biddle} in the 1D AAH chain. Such localization transition and mobility edges have also been found in coupled AAH chains~\cite{rossi}. Recently, 1D AAH 
quasi-crystal has been realized experimentally using waveguides~\cite{kraus} by Kraus {\it et al}. The authors in their work showed that the edge states of the fabricated photonic quasi-crystal are topologically nontrivial. Owing to such striking properties and several others, this model has been investigated widely in many contexts over more than three decades~\cite{ganeshprl110,ganeshprl114,ganeshprb91,harter,purka,longhi,wang,yoo,lin}.

Recently, it has been shown that {\it various topological states that are difficult to observe in condensed matter experiments, can be simulated with electric circuits}~\cite{thomale1,thomale2,helbig}. Even though electric circuits represent classical systems, with a proper choice of the reactive elements, the corresponding admittance matrix becomes equivalent to the tight-binding Hamiltonian~\cite{thomale1,thomale2,helbig,jalil-com-phys,broy}. By means of circuits, the energetics and topological phases of various physical systems have been investigated in recent years, such as Su-Schrieffer-Heeger (SSH) model~\cite{thomale2}, Weyl semimetal~\cite{jalil-com-phys}, Chern and quantum spin Hall insulators~\cite{broy}, topological Anderson insulators~\cite{zqzhang}, breathing kagome and pyrochlore lattices~\cite{ezawaprb98}, and many others~\cite{zhao,nakata,imhof,haenel,zhu,song,jalilprb}. Most of the aforementioned works focused on the study of topological phases based on the close correspondence between such electric circuits and TB models. The motivation of this work is twofold -- simulation of a TB AAH system using an electrical circuit and inspection of localization phenomena. 

\begin{figure}[h]
\centering
\includegraphics[width=0.46\textwidth]{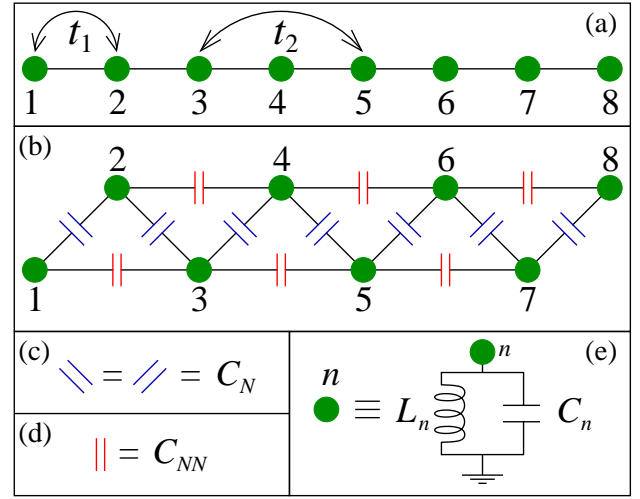}
\caption{(Color online.) (a) Schematic diagram of a 1D quasiperiodic TB chain consisting of eight sites. $t_1$ and $t_2$ are the nearest-neighbor and next nearest-neighbor hopping integrals, respectively. (b) Schematic circuit analogous to 1D quasiperiodic TB chain. (c) Blue capacitors $\left(C_N\text{'s}\right)$ connect the nearest-neighbor nodes. (d) Red capacitors $\left(C_{NN}\text{'s}\right)$ connect the next nearest-neighbor nodes. (e) The grounded connection of each node $n$.}
\label{setup}
\end{figure}

First, we construct electric circuits comprising inductors and capacitors (see Fig.~\ref{setup}) that describe TB AAH systems. The cosine modulation of usual AAH TB site energies is incorporated into the circuit by connecting different values of capacitors at the nodes of the circuit.
Primarily, we focus on designing circuits that are analogous to the TB 1D chains with nearest-neighbor (NN) and next nearest-neighbor (NNN) connections and then a two-stranded ladder network. We detect any localization behavior present in those circuits by computing a two-port impedance (TPI)~\cite{2port}.

The notable features of this Letter are: (i) realization of correlated disordered systems with electrical components, (ii) direct mapping of NN and NNN AAH TB chains and two-stranded AAH ladder with electrical circuits, and (iii) exact correspondence of admittance spectra of electrical circuits with energy spectra of TB AAH systems. Our analysis can be implemented to any other such fascinating correlated systems.


The TB Hamiltonian modeled on a 1D chain (Fig.~\ref{setup}(a)) within a non-interacting electron picture considering both NN and NNN hoppings can be written as~\cite{rossi},
\begin{equation}
H= \sum\limits_{n} \epsilon_n c_n^{\dagger} c_n +
t_{1}\sum\limits_{\langle nm\rangle}c_n^{\dagger} c_m + t_{2}\sum\limits_{\langle\langle nm\rangle\rangle}c_n^{\dagger} c_m. 
\label{ham}
\end{equation} 
Here $t_1$ is the NN hopping integral and $\langle \rangle$ represents the NN sites of the 1D chain, while $t_2$ is the NNN hopping strength and $\langle \langle \rangle \rangle$ denotes the NNN sites of the 1D chain. 
$\epsilon_n$ is the on-site potential at 
site $n$. The AAH disorder is introduced through the on-site potential and it is~\cite{aah2}
\begin{equation}
\epsilon_n=W\cos{(2\pi b n+\phi_\nu)}
\label{aah}
\end{equation}
where $W$ is the modulation strength, $b$ is an irrational number and it is chosen as $b=(\sqrt{5}-1)/2$, $n$ is the site index, and $\phi_\nu$ is the AAH phase factor.

To map the TB 1D lattice model (Fig.~\ref{setup}(a)) that satisfies Eq.~\ref{ham}, we design an electrical circuit which is given in Fig.~\ref{setup}(b). Kirchhoff's law at node $n$ in the given LC circuit reads as~\cite{broy}
\begin{equation}
\dot{I}_n = \sum_m C_{nm}\left(\ddot{V}_n - \ddot{V}_m\right) + C_n \ddot{V}_n + \frac{1}{L_n}V_n.
\label{curr}
\end{equation}
Here $\dot{V}=\frac{dV}{dt}$. $I_n$ and $V_n$ are the current and voltage at node $n$, respectively. The sum over $m$ is taken for the first and second nearest-neighbor nodes. $C_{nm}$ is the capacitor connected between nodes $n$ and $m$. In the second term of Eq.~\ref{curr}, the capacitor $C_n$ and the inductor $L_n$ (in the third term) are connected between the node $n$ and ground. 

Following the Fourier transformation of Eq.~\ref{curr}, the relationship between the current and voltage at frequency $\omega$ becomes
\begin{equation}
I_n(\omega) = \sum_m J_{nm}(\omega) V_m(\omega).
\end{equation}
Here $J_{nm}$ is known as the admittance matrix and it becomes
\begin{equation}
J_{nm} = j\omega \left[C_{nm} + \left(C_n + \left( \sum_m C_{nm} - \frac{1}{\omega^2 L_n}\right)\right)\delta_{nm}\right].
\label{admit}
\end{equation}
A one-to-one correspondence can be established between the admittance matrix $J_{nm}$ and the TB Hamiltonian (Eq.~\ref{ham}) with
\begin{eqnarray}
t_{1(2)} &=& j\omega C_{nm} \\ \epsilon_n &=& j\omega\left(C_n + \left( \sum_m C_{nm} - \frac{1}{\omega^2 L_n}\right)\right)\label{ons}.
\end{eqnarray}
Apart from the term $j\omega$, the NN hopping integral $t_{1}$ can be identified with the capacitor $C_N$ (Fig.~\ref{setup}(b)) and the NNN hopping integral $t_{2}$ with $C_{NN}$ (Fig.~\ref{setup}(c)). Setting the frequency of the input as $\omega = 1/\sqrt{L_n\sum_m C_{nm}}$, the term $\left( \sum_m C_{nm} - \frac{1}{\omega^2 L_n}\right)$ in Eq.~\ref{ons} becomes zero. Then the capacitor $C_n$ (connected between the $n$-th node and the ground) can be used to incorporate the AAH disorder into the admittance matrix.

To study the localization behavior of the circuit, the simplest experimentally measurable quantity is the two-port impedance $Z_{nm}$ between nodes $n$ and $m$. $Z_{nm}$ is defined as~\cite{2port}
\begin{equation}
Z_{nm} = \frac{V_n - V_m}{I}
\end{equation} 
where $V_n - V_m$ is the voltage difference between the nodes $n$ and $m$. $I$ is the magnitude of current $I = I_n = -I_m$, that is the current $I$ flows into node-$n$ and leaves node-$m$. 

In order to determine $Z_{nm}$, we need to express the potentials in terms of the input current $I$ and for that, the admittance matrix (Eq.~\ref{admit}) has to be inverted. To do so, first, we write the spectral form of $J_{nm}$ as
\begin{equation}
J_{nm} = \sum_p j_p \psi^*_{p,n}\psi_{p,m}
\end{equation}
where $j_p$ is the $p$-th eigenvalue of the admittance matrix and $\psi_{p,n}$ is the $p$-th eigenfunction at node $n$. With this, the regularized inverse of the admittance matrix, known as the circuit Green's function, can be written as
\begin{equation}
G_{nm} = \sum_{p,j_p\neq 0} j^{-1}_p \psi^*_{p,n}\psi_{p,m}.
\end{equation}
The two-port impedance then simplifies to
\begin{eqnarray}
Z_{nm} &=& G_{nn} + G_{mm} - G_{nm} - G_{mn}\nonumber\\ &=& \sum_p j^{-1}_p \lvert\psi_{p,n}-\psi_{p,m}\rvert^2.
\end{eqnarray}

First, we consider the circuit which is analogous to a 1D TB NN AAH chain. For this case, we do not consider any connection with the capacitors $C_{NN}$'s. Before we discuss the results, let us specify the values of the capacitors and inductors. We consider the number of nodes in the circuit as $N=100$ and set $C_N=1\,\mu$F. The inductor $L_n$ that is connected between $n$-th node and ground is set at $L_n=L=1\,$mH except at the extreme two nodes, namely nodes 1 and 100. The inductors between the ground and these two nodes are fixed at $L_1=L_{100}=2L$. The frequency of the input is fixed at $\omega=1/\sqrt{2LC_N}\, (i\neq 1,100)$. The capacitor $C_n$ (Fig.~\ref{setup}(e)) is considered in the $\mu$F range and the magnitude is chosen according to Eq.~\ref{aah}. For instance, the grounded capacitors for a disorder strength $W=1$ (with $\phi_\nu=0$) assume the values, $C_1= -0.7374\,\mu$F, $C_2= 0.0874\,\mu$F, $C_3= 0.6084\,\mu$F, and so on. In order to observe the localization behavior in the circuit, the two-port impedance is the measurable quantity. Now, from the mathematical expression of two-port impedance, one sees that the term in the denominator is the eigenvalues of the admittance matrix. If any one of the eigenvalues becomes close to zero, then the two-port impedance is ought to be very large, irrespective of the nature of the corresponding eigenstate. Thus, to extract the actual nature of a state, that is whether the state is localized or not, we attach another capacitor $C_{\text{offset}}$ (not shown in Fig.~\ref{setup}) between each node and the ground, in parallel to $C_n$. 
\begin{figure}[h] 
\centering
\includegraphics[width=0.5\textwidth]{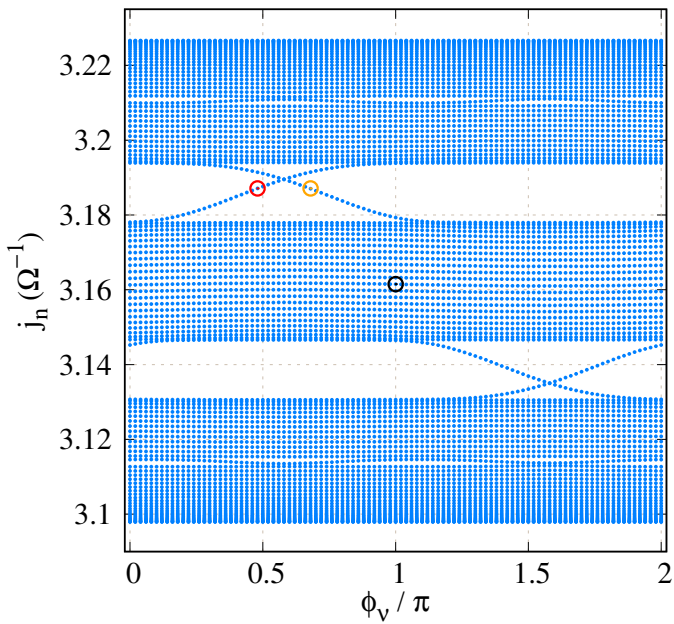}
\caption{(Color online.) Admittance spectrum as a function of Aubry phase $\phi_\nu$ considering NN connection. Number of nodes $N=100$. The AAH disorder strength is fixed at $W=0.5$. Other circuit parameters are described in the text. The nodes within the red, orange, and black circles are picked up to compute the two-port impedance in Fig.~\ref{node}.}
\label{spec-phinu}
\end{figure}
This simply shifts all the eigenvalues well above the zero line and hence makes it possible to get the true localized behavior of the eigenstates. We set $C_{\text{offset}}=10\,$mF. The effect of such a connection can readily be observed in the admittance spectra as shown in Fig.~\ref{spec-phinu}.

In Fig.~\ref{spec-phinu} we plot the eigenvalues $j_n$ of the admittance matrix as a function of AAH phase $\phi_\nu$. As mentioned earlier, the structures of the admittance and the TB Hamiltonian matrices are identical, apart from the term $j\omega$ (Eq.~\ref{admit}). The eigenvalues $j_n$ of the admittance matrix are expressed in units of $\Omega^{-1}$ and $\phi_\nu$ in units of $\pi$. The modulation strength is fixed at $W=0.5$. As mentioned above, $C_{\text{offset}}$ shifts all the eigenvalues towards a positive value and none of the eigenvalues is close to zero. The spectrum is divided into three branches and they are almost constant with $\phi_\nu$. A few modes are also seen to cross from one branch to another through the gaps with $\phi_\nu$. Overall, the behavior of the admittance spectrum as a function of $\phi_\nu$ is identical to the energy spectrum as a function of $\phi_\nu$, computed for 1D tight-binding Aubry chain by Kraus {\it et al}.~\cite{kraus}, and thus, we can claim that our circuit setup is correct.

\begin{figure}[h]
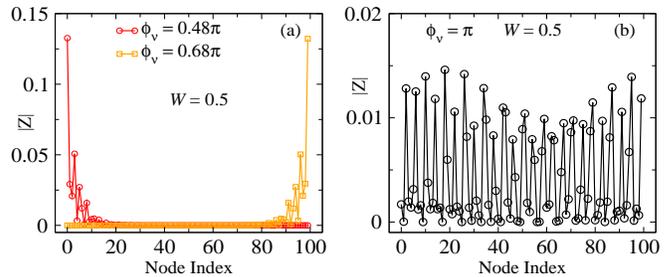

\includegraphics[width=0.235\textwidth,height=0.2\textwidth]{fig3a.eps}
\hfill
\includegraphics[width=0.235\textwidth,height=0.2\textwidth]{fig3b.eps}
\caption{(Color online.) Visualizing edge states. Two port impedance $|Z|$ as a function of node index. One port is fixed at node 50. (a) The red curve is for $\phi_\nu=0.48\pi$ and the considered eigenmode is chosen from Fig~\ref{spec-phinu}, marked with red circle. The orange curve is for $\phi_\nu=0.68\pi$ and the considered eigenmode is chosen from Fig~\ref{spec-phinu}, marked with orange circle. Both modes have the same eigenvalue. (b) The black curve is for $\phi_\nu=0.48\pi$ and the considered eigenmode is chosen from Fig~\ref{spec-phinu}, marked with black circle.}
\label{node}
\end{figure}

Now, let us look at the localization behavior of the circuit. Here we consider two particular modes in the gap, namely the modes at $\phi_\nu=0.48\pi$ (the red hollow circle in Fig.~\ref{spec-phinu}) and $\phi_\nu=0.68\pi$ (the orange hollow circle in Fig.~\ref{spec-phinu}), having the same eigenvalues. We compute the two-port impedance for these two modes as a function of node index as shown in Fig.~\ref{node}(a). 
We fixed one port at node 50 and the other one is taken through all the nodes in order to compute the two-port impedance. The results for $\phi_\nu=0.48\pi$ and $\phi_\nu=0.68\pi$ are displayed in red and orange colors, respectively. In Fig.~\ref{node}(a), we see that the two-port impedance $\lvert Z \rvert$ is maximum at the extreme left node, namely node 1 for $\phi_\nu=0.48\pi$ and abruptly decreases to zero from the left side to the right of the 1D electric circuit. On the other hand, for $\phi_\nu=0.68\pi$, we observe a complete mirror-symmetric feature of the previous case. The maximum impedance is now observed at node 100, which is at the extreme right of the 1D circuit and then gradually decreases from the right side to the left. We also choose a mode from the bulk band as shown by the black circle in Fig.~\ref{spec-phinu} and computed the two-port impedance. The corresponding result is shown in Fig.~\ref{node}(b). In the given case, $\lvert Z \rvert$ is about an order of magnitude less than the previous two cases, indicating that all the nodes are well extended for the chosen bulk mode. Overall, the behavior of two-port impedance with node index is very much consistent with the established localization behavior of 1D TB AAH chain~\cite{kraus} as a function of $\phi_\nu$. 

\begin{figure}[h] 
\centering
\includegraphics[width=0.5\textwidth]{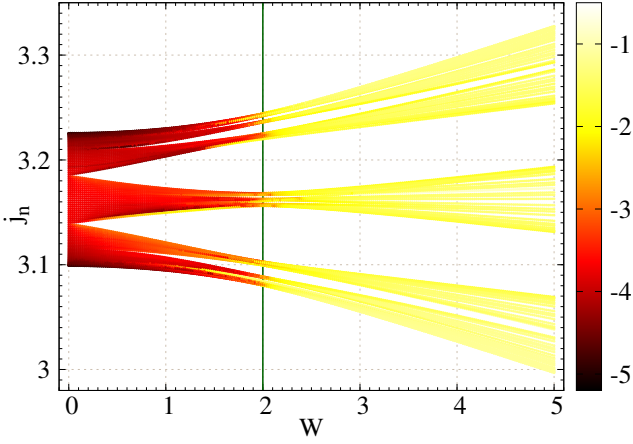}
\caption{(Color online.) Density plot of the natural log of two-port impedance as functions of modulation strength $W$ and eigenvalues of admittance matrix $j_n$, considering NN connection. Number of nodes $N=100$. The colorbar denotes the values of $\ln{\lvert Z\rvert}$. The green vertical line at $W=2C_N$ denotes the sharp transition from the highly conducting region to the low conducting one. Thus the critical $W$ is $W_c=2C_N$.}
\label{spec-dis-nn}
\end{figure}

Next, we vary the disorder strength via the grounded capacitors $C_n$s and study the behavior of two-port impedance. Here we set $\phi_\nu=0$ and all other circuit parameters are taken same as mentioned earlier. In the density plot of Fig.~\ref{spec-dis-nn}, we show the behavior of the natural log of two-port impedance as functions of disorder strength and eigenvalues of admittance matrix (measured in $\Omega^{-1}$). Here the two-port impedance 
at each of the eigenmodes is computed by keeping one port at node-1 and varying the other port at all the other nodes and then taking the maximum of $\lvert Z \rvert$. The values in the colorbar denote the natural log of two-port impedance, where lower values are denoted with dark color and higher values with bright ones. The eigenspectrum of the admittance matrix is divided into three branches as expected. Below $W=2C_N$, the computed two-port impedance for all the modes is vanishingly small, as is clearly seen in Fig.~\ref{spec-dis-nn}. Beyond $W=2C_N$, the eigenvalue spectrum becomes brighter, indicating that the TPI for all the modes is much higher than that in $W<2C_N$ region. Therefore, a sharp transition occurs at $W=2C_N$ (critical $W$ is $W_c=2C_N$) from a highly conducting zone (vanishingly small $\lvert Z \rvert$) to a low conducting one (relatively large $\lvert Z \rvert$). Such a sharp transition in the context of localization has already been studied in 1D TB AAH chain~\cite{ganeshprl114}.

\begin{figure}[h] 
\centering
\includegraphics[width=0.5\textwidth]{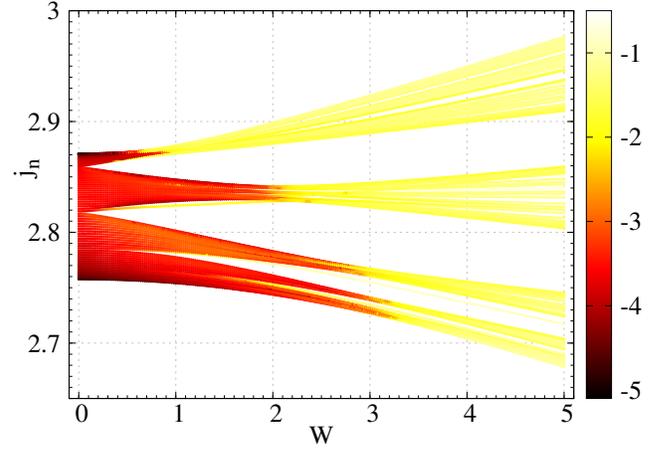}
\caption{(Color online.) Density plot of the natural log of two-port impedance as functions of  modulation strength $W$ and eigenvalues of admittance matrix $j_n$, considering NNN connection. Number of nodes $N=100$. The colorbar denotes the values of $\ln{\lvert Z\rvert}$.}
\label{nnn}
\end{figure}

Now, we bring in the capacitor $C_{NN}$ to make the circuit analogous to a 1D NNN TB AAH chain. We fix the capacitor $C_{NN}=0.25\,\mu$F. The inductor $L_n$ that is connected between $n$-th node and ground is set at $L_n=L=1\,$mH except at the extreme four nodes, namely nodes 1, 2, 99, and 100. We fix $L_1=L_{100}=2L$ and $L_2=L_{99}=2\left(C_N + C_{NN}\right)L/\left(2C_N + C_{NN}\right)$. The frequency of the input is fixed at $\omega=1/\sqrt{2L\left(C_N+C_{NN}\right)}$. 
The rest of the circuit parameters are the same as mentioned earlier. The density plot of the TPI, shown in Fig.~\ref{nnn}, is computed following the prescription described above in Fig.~\ref{spec-dis-nn}. The spectral nature of the eigenvalues of the admittance matrix with disorder strength is identical to the behavior of the eigenvalue spectrum as a 
function of disorder strength for the 1D NNN TB AAH chain. Here also we observe a transition from a conducting region to an insulating one. However, there is no such sharp transition as in the case of NN case (Fig.~\ref{spec-dis-nn}). Rather, {\it the transition is admittance dependent}. It is important to note that in the 1D AAH tight-binding chain with higher order hopping terms, there exists energy dependent mobility edge, which separates the localized wave functions from the delocalized ones. We also have a similar situation in the present case -- {\it admittance dependent mobility edge}, which separates the highly conducting region from the low conducting one.

\begin{figure}[h]
\centering
\includegraphics[width=0.46\textwidth]{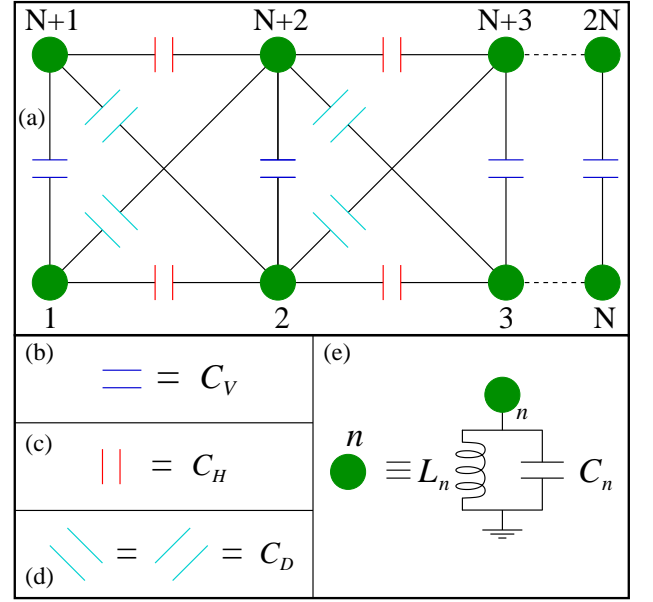}
\caption{(Color online.) (a) Schematic diagram of two-stranded ladder network having $N$ nodes in each strand. (b) Blue capacitors $C_V$'s connect the two strands vertically. (c) Red capacitors $C_H$'s connect the nearest-neighbor nodes horizontally in each strand. (d) Cyan capacitors $C_D$'s connect the nodes diagonally. (e) The grounded connection of each node $n$.}
\label{setup1}
\end{figure}

Finally, we design a two-stranded ladder network electrically. The schematic diagram for the circuit is shown in Fig.~\ref{setup1}(a).
The upper and lower strands are coupled vertically through the capacitor $C_V$ (Fig.~\ref{setup1}(b)).
In both the strands, the neighboring nodes (marked with green circles) are connected through the capacitor $C_H$, denoted with red color (Fig.~\ref{setup1}(c)).  
The crossed nodes (viz, nodes 1 and 5, nodes 2 and 4, etc.) are connected through the capacitor $C_D$ (Fig.~\ref{setup1}(d)). The $n$-th node is connected to the ground via a parallel $LC$ circuit, as shown in Fig.~\ref{setup1}(e). 

The chosen circuit parameters are as follows. The number of nodes fixed at each strand is $N=100$. The frequency of the input is fixed at $\omega=1/\sqrt{\left(2C_H+2C_D+C_V\right)L}$. The inductor $L_n$ that is connected between $n$-th node and ground is set at $L_n=L=1\,$mH ($n\neq 1,100,101,200$). The inductors between the ground and these four nodes are fixed at $L_n=\left(2C_H+2C_D+C_V\right)L/\left(C_H+C_D+C_V\right)\,(n=1,100,101,200)$.  The capacitor $C_n=C_{n+100}$ (Fig.~\ref{setup}(e)) is considered in the $\mu$F range and the magnitude is chosen according to Eq.~\ref{aah} as before. In addition to that, we attach another capacitor $C_{\text{offset}}=10\,$mF (not shown in Fig.~\ref{setup1}) between each node and ground, in parallel to $C_n$ to introduce a shift in the admittance spectrum well above the zero line due to the fact mentioned earlier.

\begin{figure}[h] 
\centering
\includegraphics[width=0.5\textwidth]{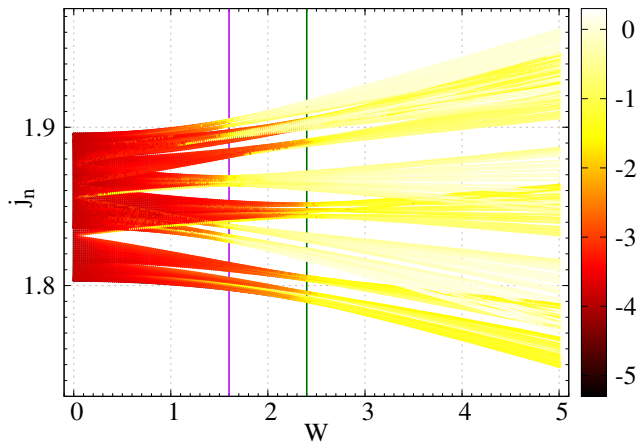}
\caption{(Color online.) Density plot of the natural log of two-port impedance as functions of modulation strength $W$ and eigenvalues of admittance matrix $j_n$ for two-stranded ladder network. Number of nodes at each strand $N=100$. The colorbar denotes the values of $\ln{\lvert Z\rvert}$. The magenta and green vertical lines denote are drawn at two critical points as described in the texts.}
\label{2strand}
\end{figure}

With all the said circuit parameters, we show the behavior of the natural log of two-port impedance as functions of disorder strength and eigenvalues of the admittance matrix in Fig.~\ref{2strand}. The computed two-port impedance for a particular eigenmode is chosen by considering all the possible two ports in the present circuit setup and then we take the maximum impedance among them. Here, the natural log of the maximum impedance is plotted for the density plot. The color convention for the colorbar is same as before. 
In the present case, we identify two critical points. The first one is $W_{c1}=2\left(C_H - C_D\right) = 1.6\,\mu$F (shown by the dark magenta vertical line), below which all the modes are highly conducting. The second one is $W_{c2}=2\left(C_H + C_D\right) = 2.4\,\mu$F, beyond which all the modes are poorly conducting in nature (shown by the dark green vertical line). Within the range $W_{c1}<W<W_{c2}$, there is a mixed phase zone, where the highly conducting modes and poorly conducting modes coexist. Such a feature is also in good agreement with the localization behavior of the two-stranded ladder network in the TB framework~\cite{rossi}.

To conclude, we have proposed a way to realize AAH disorder in electrical circuits. One-dimensional and two-stranded ladder networks have been considered for the purpose. We have shown that the AAH disorder strength and the phase can be controlled by tuning the reactive elements of the circuits. Like the inverse participation ratio (IPR), which is one of the measures of the localization phenomena in TB systems, the two-port impedance of the electrical circuits considered in the present work can serve the same purpose. Specifically, we have shown that for 1D NN circuit, the behavior of two-port impedance exhibits a sharp transition from a highly conducting region to a poorly conducting region. We have also observed admittance dependent mobility edge in 1D NNN circuit, which separates the high-conducting region from the low-conducting one. Finally, for the two-stranded ladder network, we have found two critical points, below one of the critical points, all the modes are highly conducting, and beyond the other critical point, all of them are poorly conducting. In between these two critical points, both the low and high conducting modes coexist. All the observations have been carried out based on two-port impedance. We strongly believe that the present analysis provides a direct mapping of AAH lattices with electrical circuits.

\setcounter{secnumdepth}{0}

\end{document}